# *Biosorption of Ag(I)-Spirulina platensis for different pH*

*E.Gelagutashvili, E.Ginturi, N,Kuchava, N.Bagdavadze, A.Rcheulishvili*

*Iv. Javakhishvili Tbilisi State University*
*E. Andronikashvili Institute of Physics*
*0186, 6, Tamarashvili St.,*
*Tbilisi, Georgia*

## Abstract

Biosorption of *Ag(I)-Spirulina platensis* for different *pH* were investigated using dialysis and Atomic-absorbtion analysis. It was shown, that the biosorption constant for Ag(I) *Spirulina platensis* complex and the capacity depend on the change of *pH*. In particular, with the increase of *pH (pH=5.5. and pH=8.6* cases), the biosorption constant increase and the capacity decreases. The nature of interaction is also changed. In case of neutral *pH*, the interaction Ag(I)-*S. platensis* is of cooperative character and maximum metal biosorption by *S. platensis* biomass was observed at *pH 7.0.*

## Introduction

Cyanobacteria *Spirulina platensis* (also known as blue-green algae) is gaining more attention in the field medical science because of its nutraceutical and pharmaceutical importance[1]. *S. platensis* consists of several nutritions elements, which are important for health improvement. *S. platensis* are most diverse group of photosynthetic prokaryotes. An aqueous extract of *S. platensis* inhibited HIV-1 replication in human T-cell lines, peripherial blood mononuclear cells and Langerhans cells[1].

Various species of cyanobacteria and algae have been known to adsorb and take up heavy metal ions [2]. Understending the bioavailability of heavy metals is a advantageous for plant cultivation and phytoremediation. Quantification of metal-biomass interactions is fundamental to the evaluation of potential implementation strategies, hence sorption isotherms, as well as models use to characterize algae biosorption.

The role of silver for health promotion and disease prevention is generally accepted world wide.

The aim of this study was to examine silver bioavailability with *S. platensis* for different pH.

## Materials and Methods

The study of biosorption of Ag(I)-*S. platensis* was carried out by the methods of dialysis and atomic absorption analysis. *S. platensis* was used in different state:

1. In suspension, when it is dissolved in medium and its pH=8.6.
2. Dissolved in water, pH=5.5
3. Dissolved in phosphate buffer, pH=7.0.

The experiments of dialysis were carried out in 5ml cylindrical vessels made of organic glass. A cellophane membrane of 30μm width (type - "Visking" manufacturer - "serva") was used as a partition. The duration of dialysis was 72 hours. The experiments were carried out at 20-23$^0$C temperature.

In all mentioned cases, the concentration of *Spirulina platensis* was 400mg/100ml. The metal concentration ranged from $10^{-3}$ to $10^{-5}$. The metal concentration after the dialysis was measured by atom-absorption analysis at the wavelength of $\lambda=328.1$ nm. Each value was determined as an average of three independent estimated values ± the standard deviation. Several models for the apparent Ag dissociation constant were fit to the observed binding isotherm. The best fits was obtained for binding of silver to *S.platensis* by using Freundlich and Hill models[3.4].

## Results and Discussions

Fig. 1 shows the biosorption isotherms by using the fitted Freundlich model ($LogC_b$ vs $LogC_{total}$, where $C_b$ is the ion concentration connected with cyanobacteria for all cases, and $C_{total}$ is the initial concentration of silver ions. For A and B cases, cyanobacteria were dissolved in nutrient medium *pH=8.6* and in water *pH=5.5*, respectively.

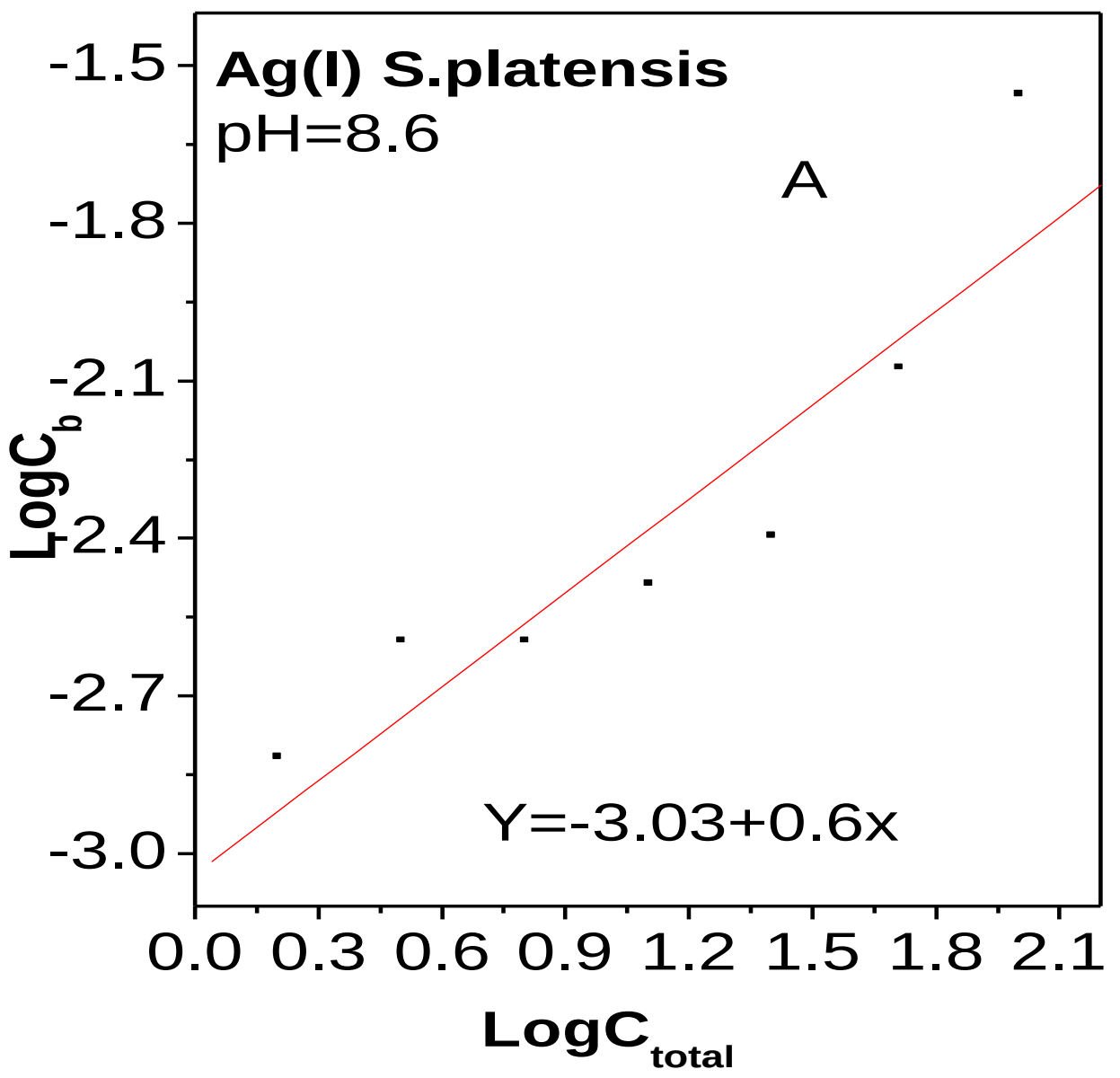

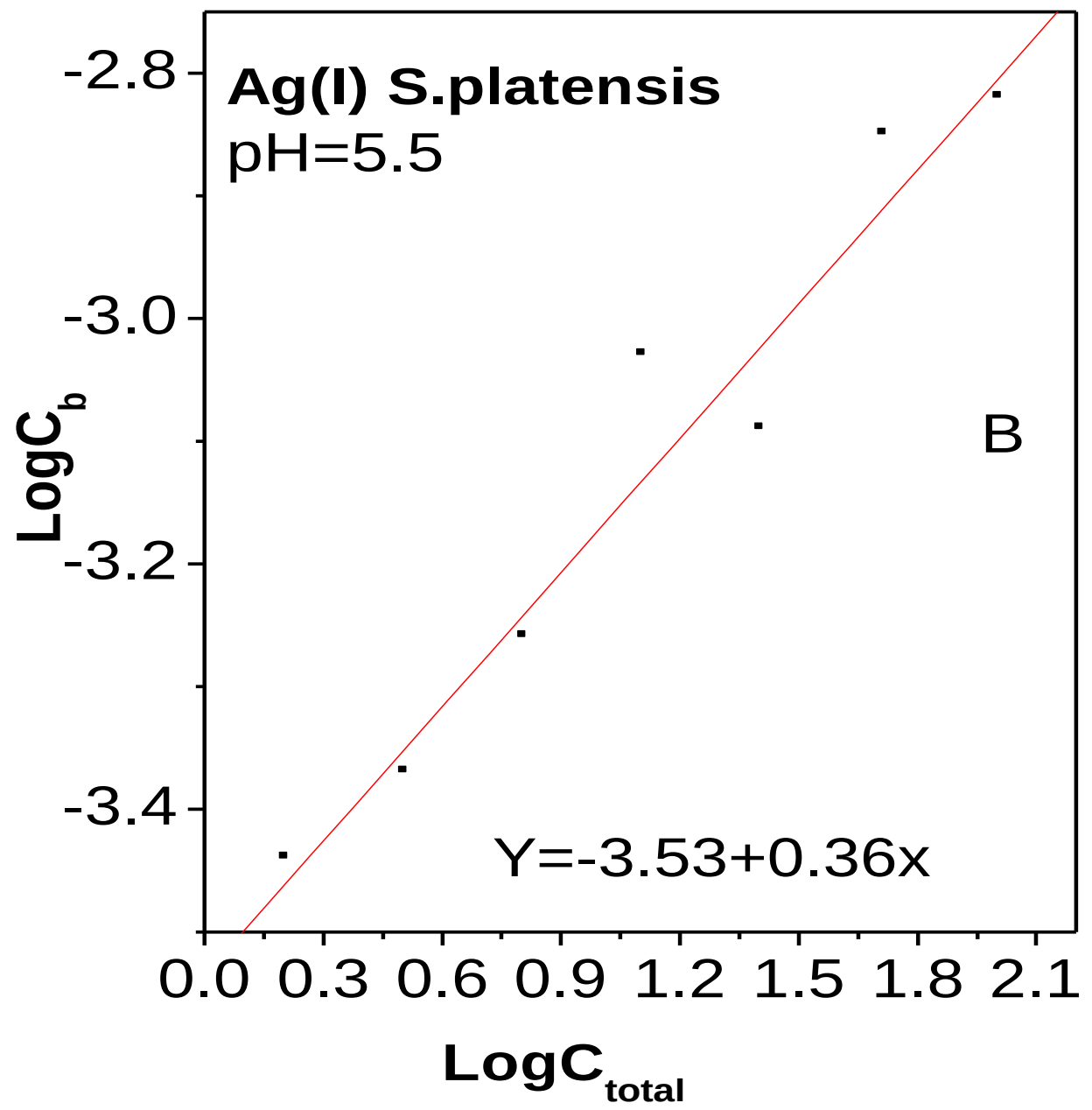

Fig. 1. Biosorption isotherms for Ag(I)-*Spirulina platensis* in the medium (A) at different *pH* and in water (B) by using the fitted Freundlich model

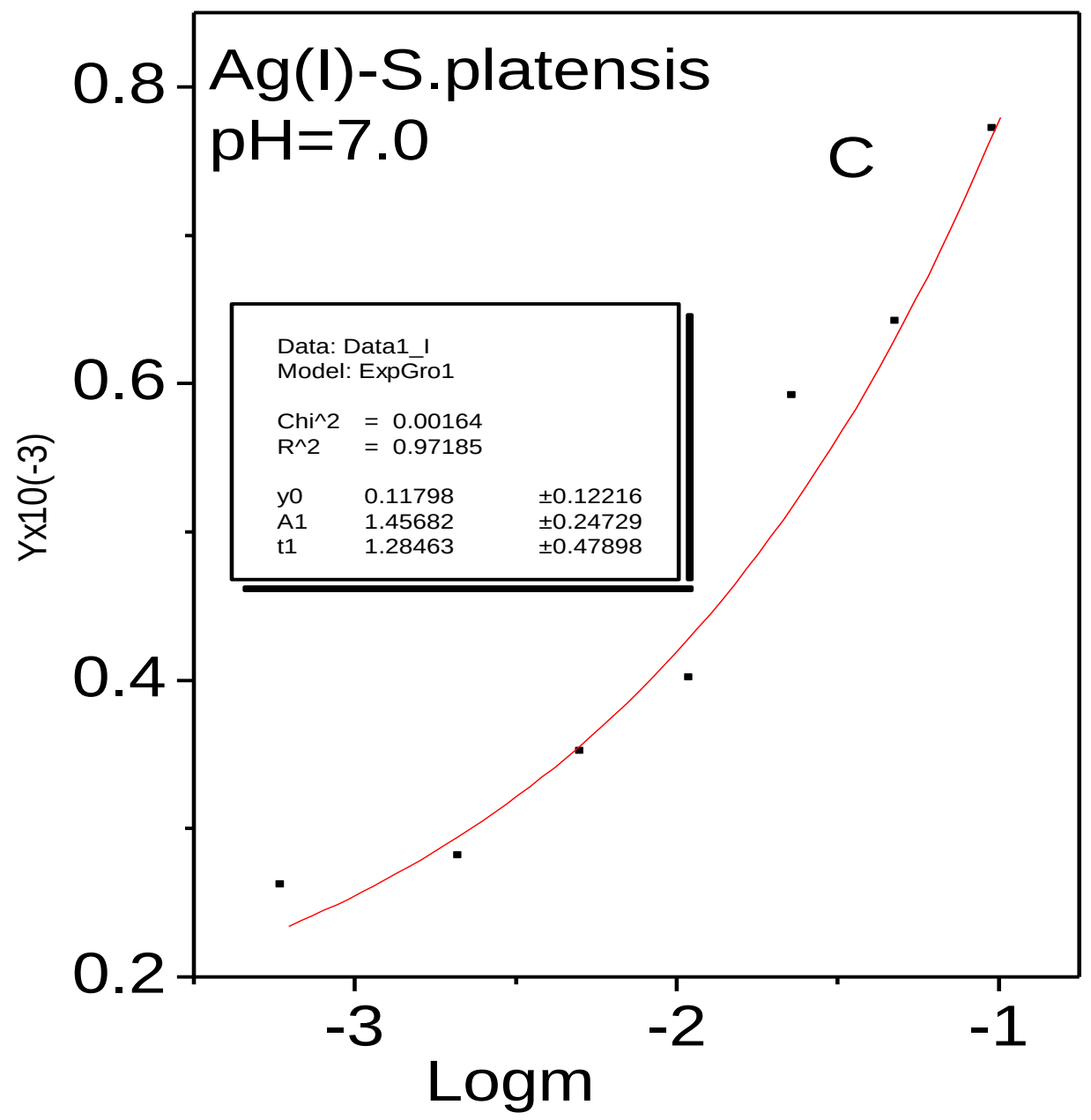

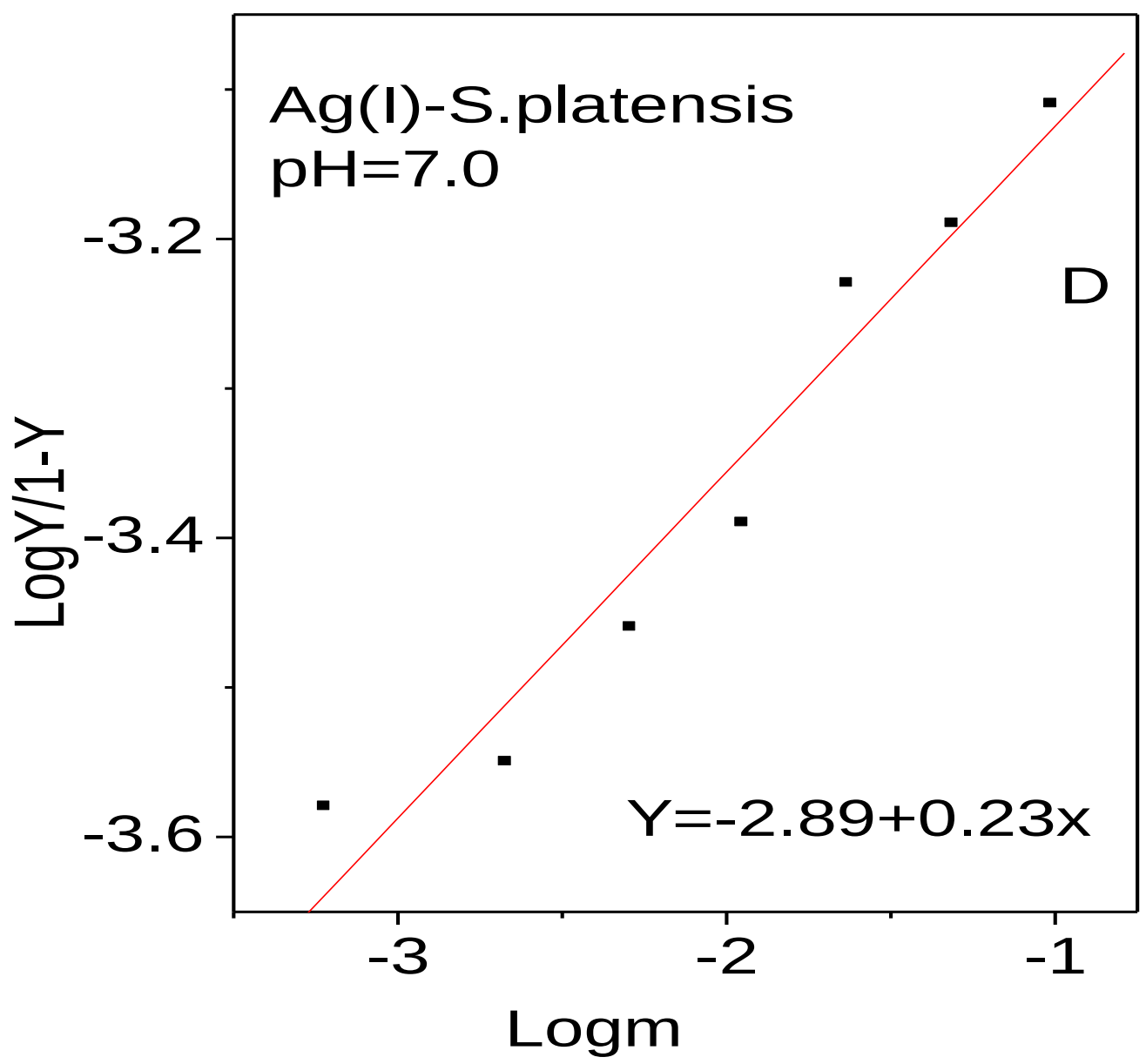

Fig. 2. Biosorption isotherms for Ag(I)-*Spirulina platensis* for neutral *pH* (dissolved in phosphate buffer).
(*C* shows the hypothetical theoretical curve chosen by $\chi^2$ criterion ($\chi^2$<0.002), *D* shows the same parameters in Hill coordinates).

By means of Freundlich isotherms the biosorption constant (*K*) and the capacity (*n*) were determined for Ag(I)-*S. platensis*. The data are shown in Table I.

Table 1. Biosorption constants for Ag(I)-Spirulina platensis

| | Biosorption constant | Biosorption capacity | Hill coefficient | Standard deviation | Correlation coefficient |
|---|---|---|---|---|---|
| | $Kx10^{-4}$ | *n* | $n_H$ | *SD* | *R* |
| *S.platensis* dissolved in medium pH=8.6 | 9.4 | 1.67 | - | 1.18 | 0.92 |
| *S.platensis* dissolved in phosphate buffer pH=7.0 | 13.0 | 5.27 | 4.34 | 0.045 | 0.97 |
| *S.platensis* dissolved in water pH=5.5 | 2.9 | 2.78 | - | 0.06 | 0.97 |

As it is seen from the Table, with the change of *pH* the biosorption parameters are changed. Namely, in case of high *pH (pH=8.6)*, the biosorption constant $K=9.4 \times 10^{-4}$ exceeds the biosorption constant $K=2.9 \times 10^{-4}$ of Ag(I)-*Spirulina platensis* for $pH=5.5$, and the capacity differs by 1.5-fold.

Fig. 2 shows the biosorption isotherm for the case, when the cyanobacteria is dissolved in phosphate buffer $pH=7.0$. The dots on the figure show the experimental data, and the solid line (C case) is the hypothetical theoretical curve chosen by $\chi^2$ criterion ($\chi^2$ 0.002 , $R^2$ 0.97) in $Y=C_{b/n}$ *vs* $Logm$ coordinates, where $n$, the number of active centers (capacity) is determined as the maximum value of $C_b$, and $m$ is the concentration of free metal ions. Each dot is the average of three independent values, and the standard deviation < 9% of average value.

The type of *y* versus *logm* dependence is nonlinear- S-shape, means that there exists positive cooperation of interaction of metal ions bound with C-PC, i.e. binding of the first metal ion increases affinity of the site for the second one. For more argumentation, as an additional criterion of cooperativity, the Hill plot (dependence *LogY/I-Y vs Logm*) was used. The line, deviation of which is the Hill coefficient ($n_H$). Taking this into account, the same data are presented in Hill coordinates, Fig. 2(D), where the parameters are the same, as in previous case, Fig. 2(C). By using Hill equality, the following values of $K$ and $n_H$ (Table 1) were obtained: $K=13.0 \times 10^{-4}$ and $n_H=4.34$. As it is seen from the Table, $n_{H/}n<1$, showing the incomplete cooperativity of the interaction between the silver ions and *S. platensis*.

In all cases, the correlation between the experimental and the theoretical data is obvious (*R* is more than *0.9*).

The mentioned model gives the possibility to characterize the interaction Ag(I)-*Spirulina platensis*. The result obtained from the cooperative interaction shows that both, the biosorption constant for Ag(I) *Spirulina platensis* complex and the capacity depend on the change of *pH*. In particular, with the increase of *pH (pH=5.5. and pH=8.6* cases), the biosorption constant increase and the capacity decreases. The nature of interaction is also changed. In case of neutral *pH*, the interaction Ag(I)-*S. platensis* is of cooperative character and maximum metal biosorption by *S. platensis* biomass was observed at *pH 7.0*.

Thus, in the case of neutral *pH*, in contrast to the previous case, it is possible to characterize the nature of interaction. In the interaction a number of functional groups take place having different location that lead to the creation of coordination centers of different type strongly differing in composition and structure. But, in the case of neutral *pH* the biosoprtion constant is the highest.